\documentclass[reprint,aps,twocolumn,superscriptaddress,floatfix,preprintnumbers,amsmath,amssymb]{revtex4-1}

\usepackage{graphics,amssymb,amsmath,epsfig,color}
\usepackage{graphicx}   
\usepackage{subfigure}
\usepackage{float}
\usepackage{dcolumn}   
\usepackage{bm}        
\usepackage{amssymb}   
\usepackage{epstopdf}
\usepackage{blindtext}

\let\oldcite=\cite 
\renewcommand{\cite}[1]{\textcolor[rgb]{0,0,1}{\oldcite{#1}}}

\let\oldref=\ref 
\renewcommand{\ref}[1]{\textcolor[rgb]{0,0,1}{\oldref{#1}}}

\begin{document}

\title{Dodecanacci superconductor-metamaterial photonic quasicrystal}

\author{Chittaranjan Nayak}
\email{83chittaranjan@gmail.com}
\affiliation{Department of Electronics and Communication Engineering, SRM Institute of Science and Technology, Chennai, 603202, Tamilnadu, India}

\author{Alireza Aghajamali}
\affiliation{Department of Physics and Astronomy, Curtin University, Perth, Western Australia 6102, Australia}

\author{Mehdi Solaimani}
\affiliation{Department of Physics, Faculty of science, Qom University of Technology, Qom, Iran}

\author{Jayanta K. Rakshit}
\affiliation{Department of Electronics and Instrumentation Engineering National Institute of technology Agartala, Agartala, India}

\author{Damodar Panigrahy}
\author{Kanaparthi V. P. Kumar}
\author{Bandaru Ramakrishna}

\affiliation{Department of Electronics and Communication Engineering, SRM Institute of Science and Technology, Chennai, 603202, Tamilnadu, India}

\begin{abstract}
Using the transfer matrix method, the present paper attempt to determine the properties of the photonic spectra of the Dodecanacci superconductor-metamaterial one-dimensional quasiperiodic multilayer. The numerical calculation is supported by using the transfer matrix method. At first, we analyze the transmission for Dodecanacci quasicrystal for different generations. After that, we analyze the effect of the thickness of the building blocks and the operating temperature. We observed that a vast number of forbidden bandgaps and transmission pecks are developed in its transmission spectra up to a certain generation number of Dodecanacci quasiperiodic sequence. If the generation number increases further, then the bandgaps become wider. According to the obtained results, depending on its generation, this structure can be used as an optical reflector or narrowband filter.
\end{abstract}

\date{\today}

\maketitle

\section{Introduction}
\label{Sec:Introduction}

Photonic crystal (PC) is a type of artificial composite material with photonic bandgap (PBG), the frequency ranges over which electromagnetic modes are forbidden, created by periodical modulated dielectric functions in the spatial domain. Since first proposed by John \cite{S_john_1987} and Yablonovitch \cite{J_E_1987}, their potential scientific and technological applications have inspired great interest among researchers. In particular, tunable PCs \cite{K_Busch_2001}, which have the capability of tuning the bandgap by external stimuli such as temperature, pressure, electric field, magnetic field, and many more, have attracted increasing interest because of their potential applications in optical devices. Therefore, the optical properties of the PCs containing various kinds of materials including dielectrics, metals, metamaterials, semiconductors, plasma, and superconductors (SCs) have been investigated \cite{J_Y_2008,A_Aghajamali_2012,A_Aghajamali_2016,J_M_2002,Z_Li_2018,A_Agha_2016_1,S_k_2016,M_Kamp_2004,H_Hojo_2004,C_Nayak_2018_1,C_Nayak_2019_1,C_Nayak_2017_1, C_Nayak_2017_2,C_Nayak_2019_2}. When the constituent materials of PCs are superconductor, or even only a defect layer is a superconductor, superconducting photonic crystals (SC-PCs) are formed \cite{C_Nayak_2019_2, S_K_2017,M_Zamani_2016,A_Agha_2016_2, A_H_2013, S_K_2014, J_Barv_2015}. The use of the superconducting photonic crystals has an advantage over conventional metal-dielectric PCs; for example, the loss issue of the PCs can be remedied by the utilization of superconductors, dielectric function of the superconductor depends on the external temperature, etc. Therefore, SC-PCs have an excellent opportunity for application in the tunable photonic crystal.

Over 30 years ago, Veselago \cite{V_G_2016} theoretically studied the propagation of electromagnetic waves in materials characterized by simultaneously having a negative permittivity ($\epsilon$) and permeability ($\mu$) and referred to such materials as left-handed materials to emphasize the fact that the electric field ($\bf{E}$), the magnetic field ($\bf{H}$) and the propagation wave vector ($\bf{k}$) are related by the left-handed rule. These materials, which are now called double-negative (DNG) materials (or metamaterial), have received extensive attention due to their experimental validation and unusual electromagnetic properties. Recently, by the possibility of manufacturing PCs with metamaterials, called metamaterial photonic crystals (MPCs), a new research area has emerged, and numerous interesting results have been reported by researchers so far. The inclusion of metamaterial in PC led to the emergence of new mechanisms to produce photonic gaps, which helps to design dielectric mirrors, effective waveguides, filters, perfect lenses, etc. Based on the properties of metamaterials, superconductor, and PCs, researchers now intend to investigate the transmittance of one-dimensional PCs in the visible and microwave regions. It was suggested that the combination of superconductor NbN with DNG metamaterials is possible to design the metamaterial superconductor PC because of the feasibility of coating \cite{A_H_2016}.

Moreover, artificially fabricated deterministic or so-called quasiperiodic structures constitute a separate field of research. These quasiperiodic photonic structures are formed by the superposition of two (or more) incommensurate periods so that they can be defined as intermediate systems between a photonic crystal and the random photonic multilayer. Compared to the periodic and random photonic multilayers, photonic quasicrystals (PQCs) share distinctive physical properties with both periodic media, i.e., the formation of well-defined PBGs, and disordered random media, i.e., the presence of localized states, thus offering an almost unexplored potential for the control and manipulation of localized field states. This conception of quasicrystal was first introduced by Kohmoto~\emph{et al.} \cite{M_Kohmoto_1987} in the field photonics with a Fibonacci arrangement of two dielectric materials. Then after various quasicrystals including, Thue-Morse \cite{C_Nayak_2019_1, C_H_2017,H_Rahimi_2019}, double-periodic \cite{C_Nayak_2019_1, H_Rahimi_2019}, Rudin-Shapiro \cite{ H_Rahimi_2019}, Octonocci \cite{C_Nayak_2019_1, M_Zamani_2019,C_nayak_2017_3}, and many more are presented with interesting and useful results. Brief reports on this topic were presented by Bellingeri \cite{ M_Belli_2017}, Vardeny~\emph{et al.} \cite{ Z_V_2013}, and Edagawa \cite{ K_Edagawa_2014} in which various PQCs structures were discussed. Recently it was presented that the cutoff frequency and the bandwidth of superconducting photonic quasicrystals (SC-PQCs) are remarkably sensitive to the temperature of superconducting material, periodic short order of the quasiperiodic system, incident angles, and type of polarization \cite{ H_Rahimi_2019, Y_Trabelsi_2019, R_Talebzadeh_2018}. The addition of metamaterial as a constituent material in the SC-PQCs is quite feasible and be an interesting topic in the field of photonic multilayer design.  

Recently, few studies on Dodecannaci photonic quasicrystals  \cite{ R_Luck_1993}  have been carried out using dielectric  \cite{ E_F_2019} and magnetized plasma  \cite{ C_nayak_2020}. Silva et al. presented the dielectric Dodecannaci \cite{ E_F_2019} with the SiO$_2$/TiO$_2$ multilayers, where a graphene monolayer at the interfaces between distinct layers. The result indicates that the whole optical spectrum becomes affected by the presence of graphene in the interfaces including the shift of bandgaps to high-frequency regions, the emergence of a graphene induced bandgap at low-frequency regions, and the decreasing of transmittance in the whole frequency range.  In another investigation, Nayak \cite{ C_nayak_2020} presented Dodecannaci extrinsic magnetized plasma quasicrystal, which evident the robust against layer position photonic bandgap aroused above the plasma frequency. Motivated by these distinctive results offered by the Dodecannaci quasicrystal and the ability to manage the electromagnetic waves by superconductor as well as the metamaterial, we now present the study of the photonic bandgap characteristics of one-dimensional Dodecannaci superconductor-metamaterial quasicrystal. 

This manuscript is organized as follows. In Section \ref{Methodology}, we describe the basic Dodecanacci quasi-sequence, the permittivity of the constituent materials, and summarize the numerical methods. In Section \ref{Result and Discussions}, we present the numerical results and discuss many physical parameters that could tune the transmittance of our proposed superconductor metamaterial quasicrystal. Finally, the conclusions are summarized in Section \ref{Conclusions}.

\section{Methodology}
\label{Methodology}

Before proceeding to the results and simulated data, in this section, we are briefly describing the fundamentals of the proposed study. This section is organized in the following three sub-sections:

\begin{itemize}
    \item Dodecanacci quasi-sequence 
    \item Constituent materials
    \item Transfer-matrix method
\end{itemize}

\subsection{Dodecanacci quasi-sequence}
\label{Dodecanacci quasi-sequence}

A generalized Dodecanacci quasi-sequence \cite{ R_Luck_1993,E_F_2019,C_nayak_2020} can be stated by the deterministic rule, which is defined as $D_x=\left\{AD_{x-2}D_{x-1}\right\}^2D_{x-1},$ for $x\geq3$, with initial condition: $D_1=AAB$, $D_2=\left\{AD_1\right\}^2D_1=AAABAAABAAB$.  Here $x$ is the generation number. The total number of blocks $A$ and $B$ in $D_x$ can be calculated by the recurrence relation, $P_x=4P_{x-1}-\ P_{x-2}$ for $x\geq3$, where $P_1=3\ and \ P_2=11$. Here, $A$ is the building block modelling of one of the constituent material having thickness, $d_A$ with permittivity $\varepsilon_A$ whereas $B$ is the building block modelling of other constituent material having thickness, $d_B$ with permittivity $\varepsilon_B$. The Dodecanacci superlattice can also be generated by the inflation rule $A\rightarrow AAAB$, and $B\rightarrow AAB$. In Table~\ref{table-1}, we illustrate the three distributed chain $D_3$, $D_4$, and $D_5$ of two building blocks $A$ and $B$ organized according to Dodecanacci quasiperiodic sequence having $x\geq3$. The other two Dodecanacci quasiperiodic sequence used in the work are $D_6$ and $D_7$ having 2131 and 7953 number of building layers, respectively. 

\newcommand\MR{\multirow{2}{*}}

\begin{table*}[!]
\caption{Organized blocks $(A,\ B)$ follow Dodecanacci quasiperiodic sequence.}
\centering
\resizebox{\textwidth}{!}{\begin{tabular}{ c c c}
\hline \hline  \\ [-2mm]
Sequence:   & Distributed ${A,B}$ chain of Dodecanacci &~~~ Total  number  \\
$D_x=\left\{AD_{x-2}D_{x-1}\right\}^2D_{x-1}$, for x$\geq$3 ~~~ &  quasiperiodic sequence  & ~~~of layers  \\  [0.9mm]
\hline \hline  \\ [-3mm]
$D_3$ &  $AAABAAABAAABAABAAABAAABAAABAABAAABAAABAAB$ & 41	\\ [1mm]
\hline \\ [-3mm]
$D_4$ &  $AAAABAAABAABAAABAAABAAABAABAAABAAABAAABAA$ & 153	\\ [1mm]
	   &  $BAAABAAABAABAAAABAAABAABAAABAAABAAABAABAA$ &  		\\ [1mm]
	   &  $ABAAABAAABAABAAABAAABAABAAABAAABAAABAABAA$ &  		\\ [1mm]
	   &  $AABAAABAAABAABAAABAAABAAB$ 					 &  		\\ [1mm]
\hline \\ [-3mm]
$D_5$ &  $AAAABAAABAAABAABAAABAAABAAABAABAAABAAABAA$ & 571	\\ [1mm]
	   &  $ABAAAABAAABAABAAABAAABAAABAABAAABAAABAAAB$ &  		\\ [1mm]
	   &  $AABAAABAAABAABAAAABAAABAAABAAABAAABAABAAA$ &  		\\ [1mm]
	   &  $ABAAABAABAAABAAABAAABAABAAABAAABAAABAABAA$ &  		\\ [1mm]
	   &  $ABAAABAABAAABAAABAAABAABAAABAAABAAABAABAA$ &  		\\ [1mm]
	   &  $AABAAABAABAAAABAAABAAABAABAAABAAABAAABAAB$ &  		\\ [1mm]
	   &  $AAABAAABAABAAAABAAABAABAAABAAABAAABAABAAA$ &  		\\ [1mm]
	   &  $BAAABAAABAABAAABAAABAABAAAABAAABAABAAABAA$ &  		\\ [1mm]
	   &  $ABAAABAABAAABAAABAAABAABAAABAAABAABAAABAA$ &  		\\ [1mm]
	   &  $ABAAABAABAAABAAABAAABAABAAABAAABAABAAAABA$ &  		\\ [1mm]
	   &  $AABAABAAABAAABAAABAABAAABAAABAAABAABAAABA$ &  		\\ [1mm]
	   &  $AABAABAAAABAAABAABAAABAAABAAABAABAAABAAAB$ &  		\\ [1mm]
	   &  $AAABAABAAABAAABAABAAABAAABAAABAABAAABAAAB$ &  		\\ [1mm]
	   &  $AAABAABAAAB$ 									 &  		\\ [1mm]
\hline
\hline
\end{tabular}}
\label{table-1}

\end{table*}

\subsection{Constituent materials}

The proposed quasicrystal composed by double-negative metamaterial (DNG) and lossless superconductor is arranged as per the Dodecanacci quasiperiodic sequence. The description of frequency dependent dielectric properties of the DNG, represents layer $A$ and the lossless superconductor represents layer $B$ in the Dodecanacci quasi sequence are appended below. The frequency dependent numerical formula for calculating the permittivity, $\varepsilon_A(f)$ and the permeability, $\mu_A(f)$  of the DNG material are given by \cite{A_Aghajamali_2012,C_Nayak_2017_2,A_H_2016}.
\begin{equation}\label{eq-1}
\varepsilon_A\left(f\right)=1+\frac{5^2}{{0.9}^2-f^2-if\gamma_e}+\frac{{10}^2}{{11.5}^2-f^2-if\gamma_e}
\end{equation}

\begin{equation}\label{eq-2}
\mu_A\left(f\right)=1+\frac{3^2}{{0.902}^2-f^2-i2\pi\omega\gamma_m}
\end{equation}
where, $\gamma_e$ and $\gamma_m$ represent electric damping frequency and magnetic damping frequency, respectively. 

The frequency dependent dielectric constant of a lossless superconductor is given by \cite{C_Nayak_2019_2,A_H_2016}
\begin{equation}\label{eq-3}
\varepsilon_B(f)=1-\left(\frac{1}{{4\pi^2f}^2\mu_0\varepsilon_0\lambda_l^2}\right)
\end{equation}
The permittivity and permeability of air are noted in the above equations with $\varepsilon_0$ and $\mu_0$ respectively. Whereas $\lambda_l$ stands for the temperature-dependent London penetration depth, which is given as
\begin{equation}\label{eq-4}
\lambda_l=\ \lambda_0/\sqrt{(1-{(T/T_c)}^4})
\end{equation}
here, London penetration depth at absolute temperature, operating temperature (K) and critical temperature (K)  is  mentioned as $\lambda_0$, $T$, and $T_c$, respectively.

\subsection{Transfer-matrix method}

To investigate the deferent aspects of  the one-dimensional Dodecannaci quasicrystal here we take the help of transfer matrix method \cite{M_Born_2013}. The characteristic matrices for layer $A$ and layer $B$ of the Dodecannaci quasicrystal is written as
\begin{equation}\label{eq-5}
 M_A  =  \left[
	\begin{array}{cc}
	\cos{\varphi_A}&\frac{-i}{p_A}\sin{\varphi_A} \\
		-ip_A\sin{\varphi_A}&\cos{\varphi_A}
	\end{array}
	\right]
\end{equation}	
\noindent and
\begin{equation}\label{eq-6}
 M_B  =  \left[
	\begin{array}{cc}
	\cos{\varphi_B}&\frac{-i}{p_A}\sin{\varphi_B}\\-ip_B\sin{\varphi_B}&\cos{\varphi_B}
	\end{array}
	\right]	
\end{equation}
\\
In Eq. \ref{eq-5}, $\varphi_A=(\omega/c)n_Ad_A\cos{\theta_A}$, c is the speed of light in vacuum, $n_A$ is the refractive index of the $A$ layer, $\theta_A$ is the ray angle inside the layer $A$, $p_A=\ \sqrt{\varepsilon_A/\mu_A}\cos{\theta_A}$ and $\cos{\theta_A}= \sqrt{1-({n_0}^2\sin^2{\theta_0/}{n_A}^2)}$, in which $n_0$ is the refractive index of the air. In Eq. \ref{eq-6} the notations are same Eq. \ref{eq-5} but for layer $B$.

The transfer matrix of the Dodecannaci quasicrystal, $M_x$ for $x\geq3$ is give as \cite{E_F_2019}
\begin{equation}\label{eq-7}
M_x=\ \left\{M_AM_{x-2}M_{x-1}\right\}^2M_{x-1}=\left[
	\begin{array}{cc}
	m_{11}&m_{12}\\m_{21}&m_{22}
	\end{array}
	\right],
\end{equation}
\\
here,  $m_{11}$, $m_{12}$, $m_{21}$ and $m_{22}$ are the matrix elements of the multilayer system $M_x $ with the initial conditions of  $M_1=M_AM_AM_B$ and $M_2=\ \left\{M_AM_1\right\}^2M_1$. The transmission coefficient $(t_D)$ of the Dodecannaci quasicrystal is given by 
\begin {equation}\label{eq-8}
t_D=\ \frac{E_O}{E_i}=\ \frac{2P_{in}}{\left(m_{11}+m_{12}P_{out}\right)P_{in}+\left(m_{21}+m_{22}P_{out}\right)}
\end{equation}
Where, $E_O$ , and $E_i$ are the input and output electric field intensity, respectively. $P_{in}=\ n_0\cos{\theta_0}$, and $P_{out}=n_{out}\cos{\theta_{out}}$ in which $n_{out}$ is the refractive index of the environment having ray angle of $\theta_{out}$. The transmittance $(T_D)$ of the Dodecannaci quasicrystal is represented by 
\begin {equation}\label{eq-9}
T_D=t_D\bullet{t_D}^\ast                               
\end{equation}                                           

\section{Result and Discussions}
\label{Result and Discussions}

We have studied in detail different realizations of the Dodecannaci superconductor-metamaterial photonic quasicrystals, composed of DNG metamaterial and low-temperature superconductor materials, NbN ($T_{C}=16$~K and $\lambda_L$(0)=200~nm) \cite{A_H_2016,W_Buckel_2004}, by varying the generation number, $x$, thickness of DNG metamaterial, $d_A$, thickness of superconductor material, $d_B$, and the operating temperature, $T$. The considered generation number,  $x$ of Dodecannaci quasicrystal, are from 1 to 7. The frequency range is fixed 1.5 to 3 GHz. The specific frequency range was chosen to fit our objective to design Dodecannaci superconductor-metamaterial photonic quasicrystals \cite{A_Agha_2014_4}. The medium surrounding the supposed structures is a vacuum. The parameters of DNG metamaterial, electric damping frequency, $\gamma_e$ and magnetic damping frequency, $\gamma_m$ are assumed to be the same and are equal to $2\ \times{10}^{-3}$  GHz \cite{A_Aghajamali_2012,C_Nayak_2019_2,A_H_2016,A_Agha_2014_4,A_Agha_2013,A_Agha_2014_5}. At this starting point, the thicknesses $d_A$ and $d_B$, and operating temperatures are set to 10 mm, 4 nm and 4.2 K as in previous reports, otherwise, it is mentioned in the respective caption of figures. 

\begin{figure}[!t]
\centering
\includegraphics[width=0.48\textwidth]{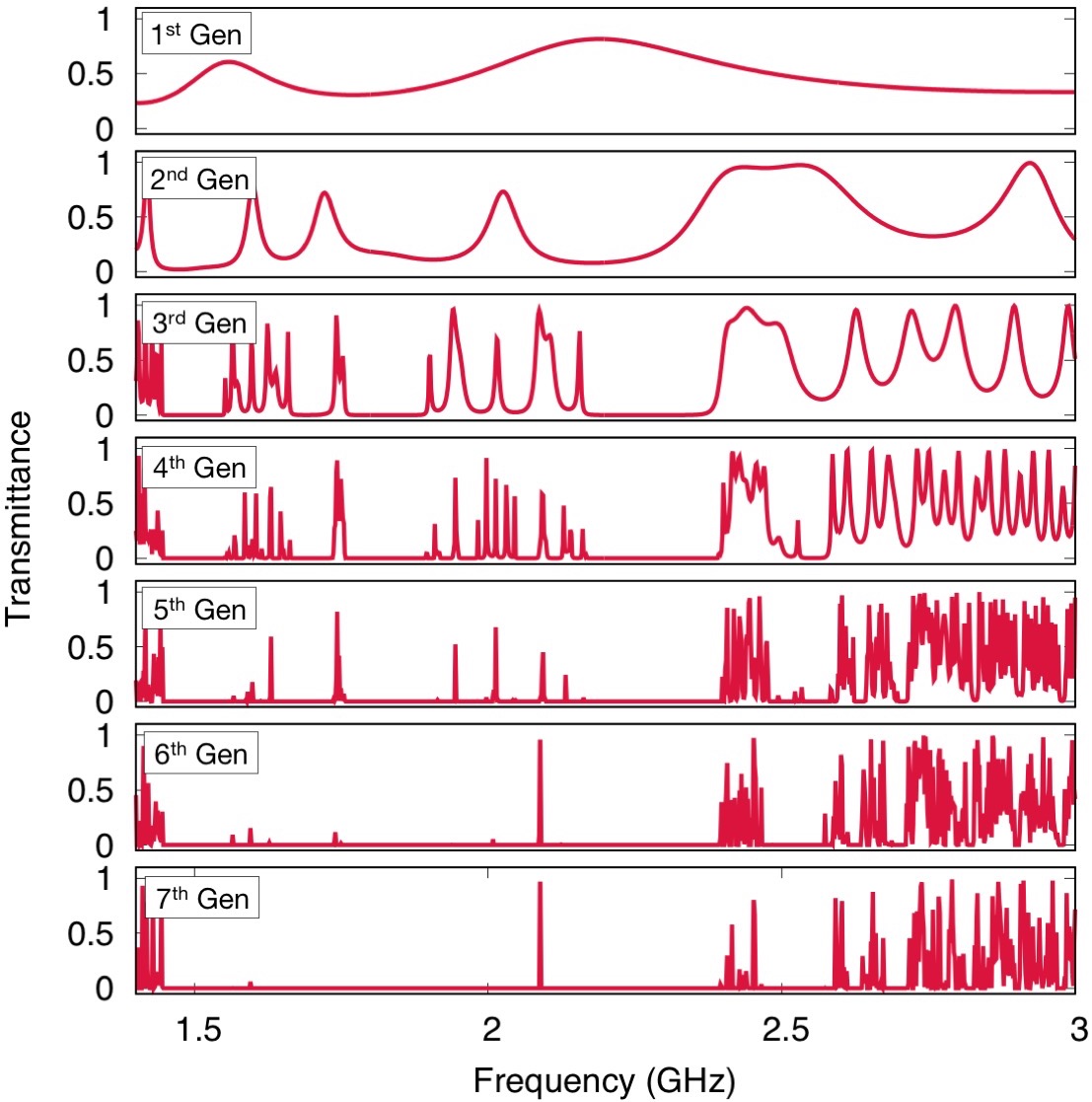}
\caption{Transmittance spectrum from 1D multilayered stack arranged according Dodecannaci sequence at generation number, $x= 1$ to $7$, while $d_A$ and $d_B$ are fixed to be 10~mm and 4~nm. }
\label{fig1}
\end{figure}

Using expressions (1) to (9) transmission spectra of Dodecannaci superconductor-metamaterial photonic quasicrystal at normal incidence is plotted in Fig.~\ref{fig1}. Here, seven sub-plots presented the transmission spectra of desired quasicrystals having for generation numbers from $x = 1$ to 7 and are clearly represented with an appropriate legend. On increasing the generation number, $x$, it is understood that the average transmission for the considered range of frequency is decreased. However, the rate of decrease in the average transmission is region-specific. More clearly, the average transmission of the first half of the considered frequency (1.5 to 2.25~GHz) is very firstly decaying as compared to the second half (2.25 to 3~GHz). This is mainly possible because of the magnitude of the negative refractive index and the number of DNG metamaterial layers in the considered stratified structures.

From the transmission Dodecannaci superconductor-metamaterial photonic quasicrystal having $x = 2$, it is clearly understood that four-measure bandgaps are appeared in between 1.5 to 2.4~GHz. These aroused bandgaps are marginally affected with an increase in $x$, whereas the transmission regions are splits and form narrow transmission regions. This trend of splitting of transmission is appeared up to $x = 4$ after that the count of the transmission peaks decrease with increase in  $x$. At, $x = 7$, we observe a very sharp transmission peak with almost unity in magnitude around 2.1~GHz, whereas a peak having nearly 5\% of transmission is also noted. These transmission peaks are because of the multiple transmission and reflection from the individual elements of the stratified structure, there spatial arrangement, and material properties. These results may be suitable for designing of narrow band filter for different microwave photonics applications.

\begin{figure}[!t]
\centering
\includegraphics[width=0.48\textwidth]{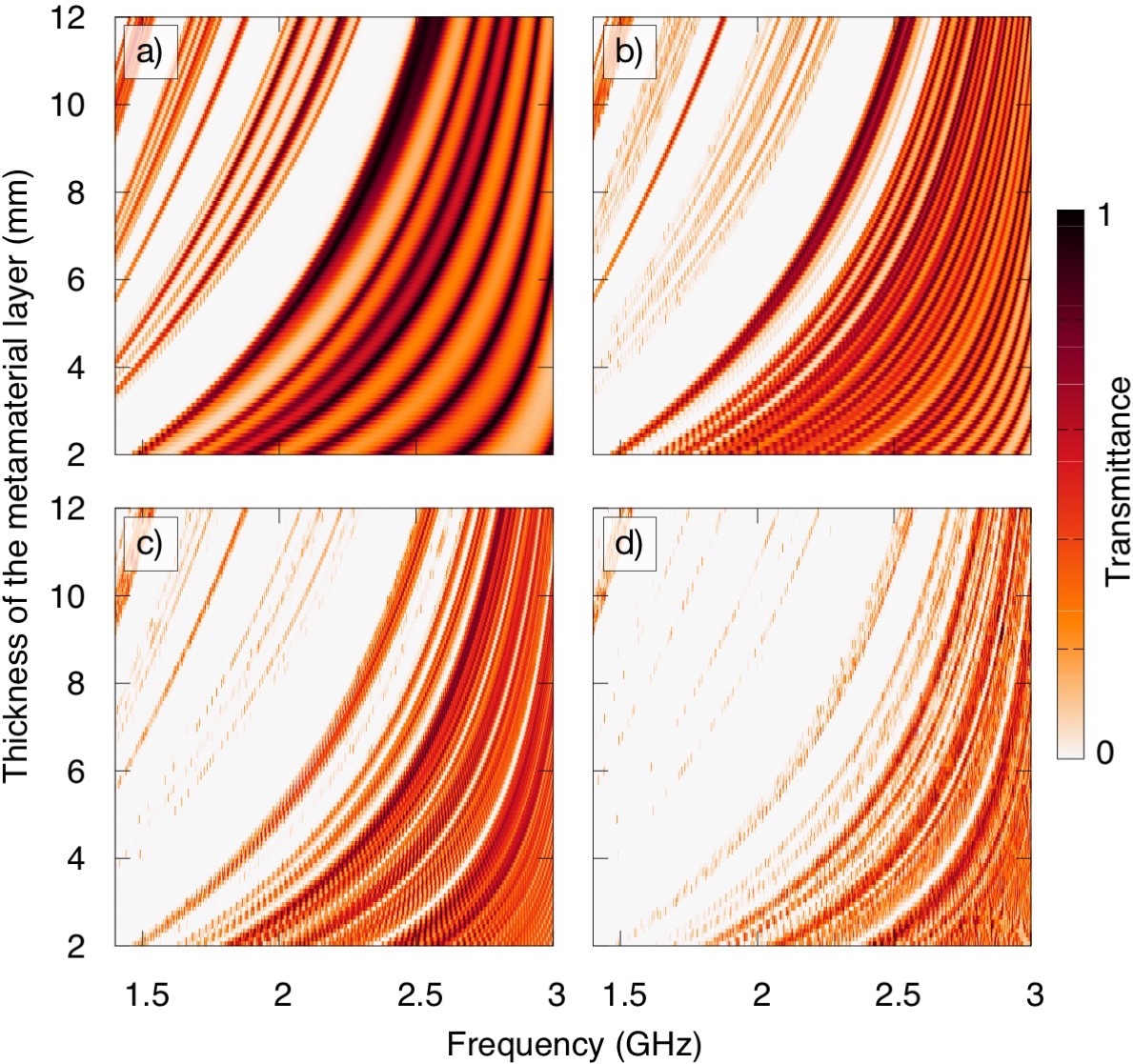}
\caption{Thickness of the DNG metamaterial dependent transmittance spectrum from 1D multilayered stack arranged according Dodecannaci sequence at generation number, $x= 3$ (a), $x= 4$ (b), $x= 5$ (c), and  $x= 7 $ (d) while $d_B$ is fixed to be 4~nm.}
\label{fig2}
\end{figure}

As we know the transmission spectra is a function of the thickness of building blocks, here the investigation is performed over a range of thicknesses of DNG metamaterial, $d_A$ as well as thicknesses of superconductor material, $d_B$. Fig.~\ref{fig2} represents the thickness of the DNG metamaterial dependent transmittance spectrum from 1D multilayered stack arranged according to Dodecannaci sequence at generation number, $x = 3$ (a), $x = 4$ (b), $x = 5$ (c), and  $x = 7$ (d). Here, excluding $d_A$ all other parameters that are used to compute the transmission spectra in Fig. 1 are not changed. By observing the results, it is noted that the transmission spectra are the function of the thicknesses of the DNG metamaterial, $d_A$ and blue shifted. This observation is obvious and resemblance to the previous findings \cite{A_H_2016}. However, it is also interesting to note that the transmission peaks have become more extensive as the thicknesses of the DNG metamaterial, $d_A$ increases. 

\begin{figure}[!t]
\centering
\includegraphics[width=0.48\textwidth]{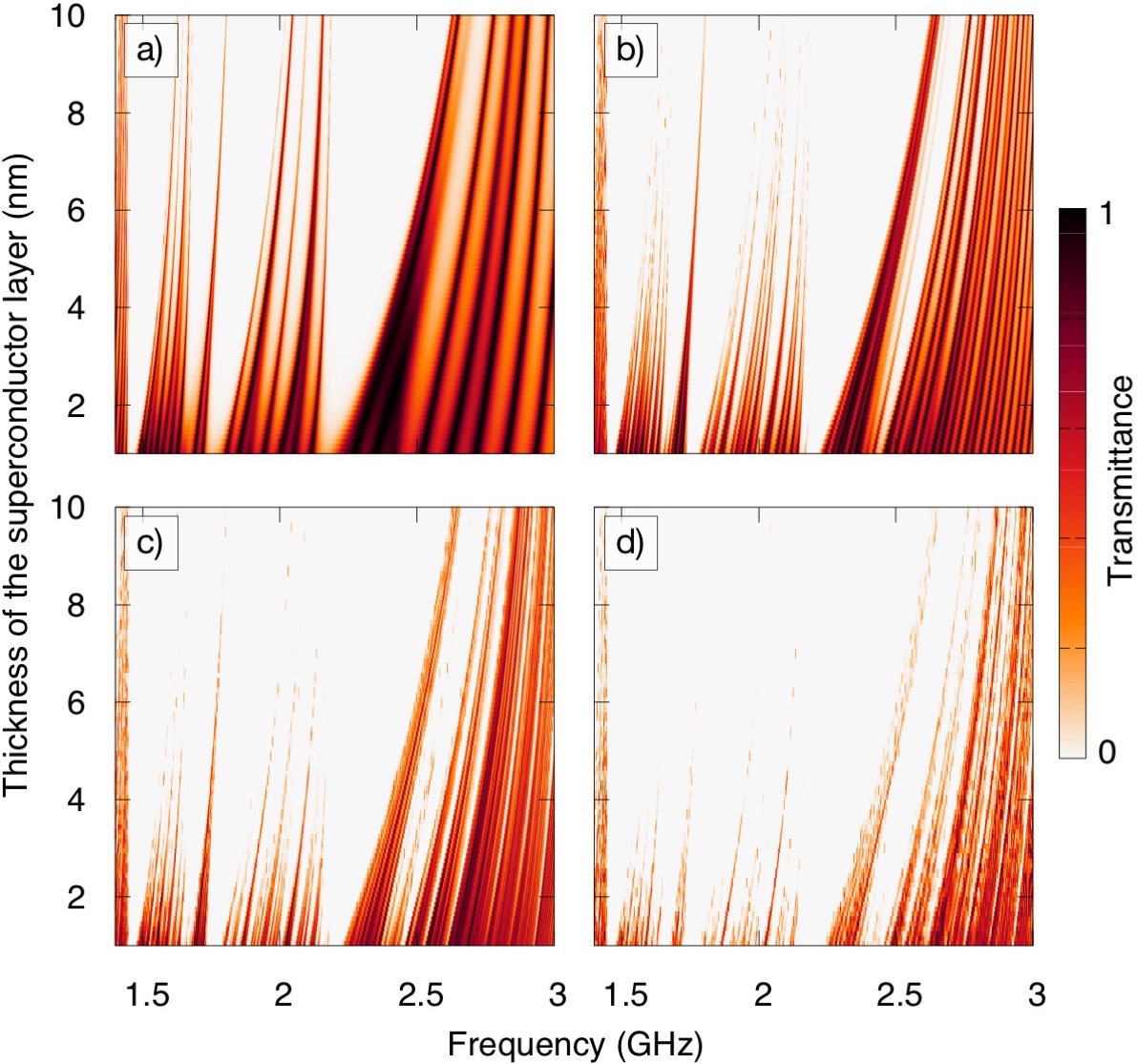}
\caption{Thickness of the superconductor dependent transmittance spectrum from 1D multilayered stack arranged according Dodecannaci sequence at generation number, $x= 3$ (a), $x= 4$ (b), $x= 5$ (c), and  $x= 7 $ (d) while $d_A$ is fixed to be 10~mm.}
\label{fig3}
\end{figure}

In Fig.~\ref{fig3} we present the thickness of the superconductor layers dependent transmittance spectrum from multilayered stack arranged according Dodecannaci sequence at generation number, $x= 3$ (a), $x= 4$ (b), $x= 5$ (c), and  $x= 7$(d). As like our first case of thickness variation for DNG metamaterial, here, excluding $d_B$ all other parameters that are used to compute the transmission spectra in Fig.~\ref{fig1} are not changed. By observing the results, it is noted that the response of the shifting of transmittance spectra is similar to the change in layer thickness DNG metamaterial, $d_A$ but with a reduced rate of blue shift. While observing the width of the transmission peak, here we got an inverse effect $i.e.$, the width of the transmission peak is decrease with an increase in the thickness of the superconductor layers. The physics behind these observed results is the effective optical path length caused by the constituent materials of the proposed supper lattice. 

\begin{figure}[!t]
\centering
\includegraphics[width=0.48\textwidth]{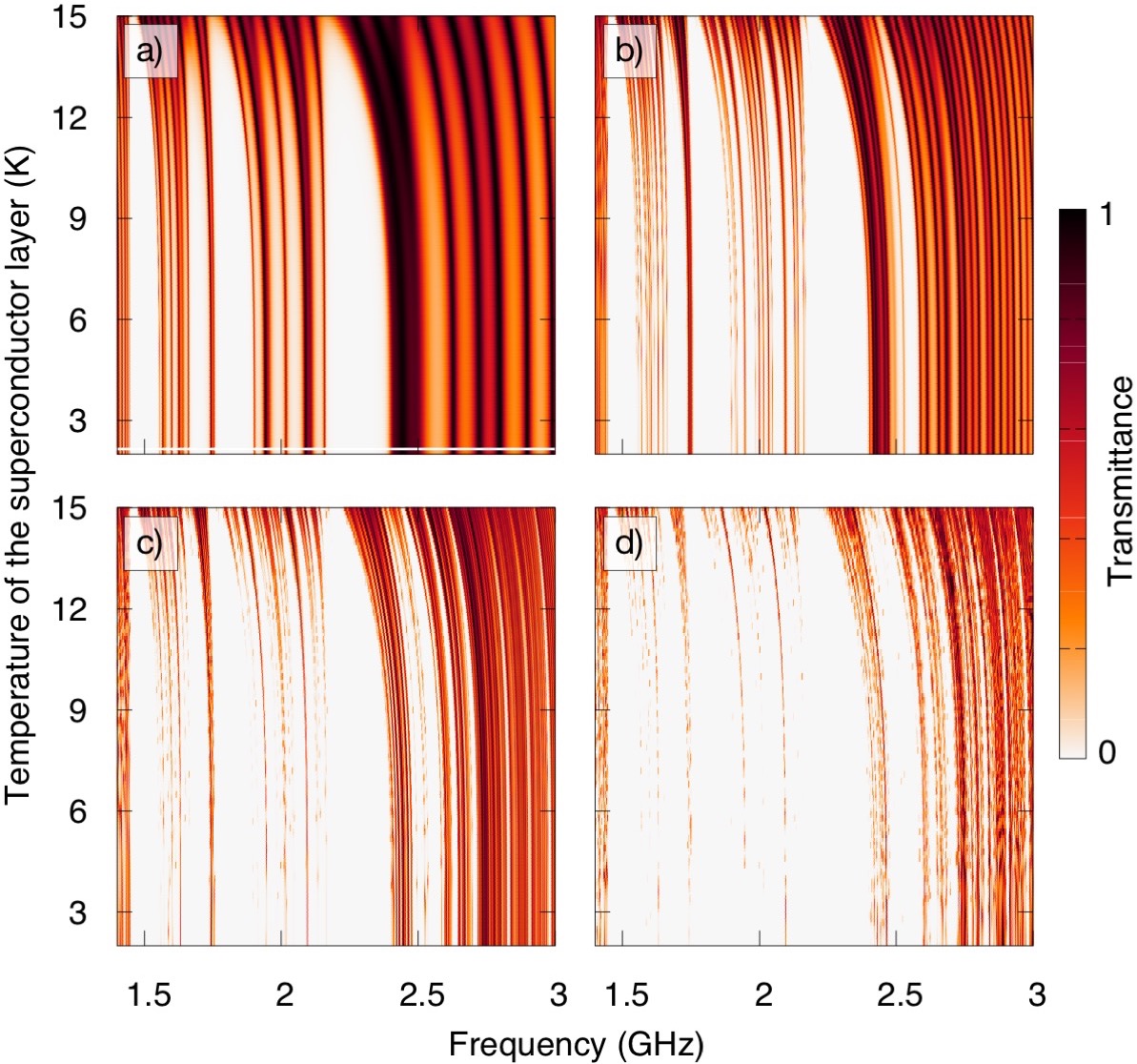}
\caption{Temperature dependent transmittance spectrum from 1D multilayered stack arranged according Dodecannaci sequence at generation number,  $x= 3$ (a), $x= 4$ (b), $x= 5$ (c), and  $x= 7 $ (d) while $d_A$ and $d_B$ are fixed to be 10~mm and 4~nm.}
\label{fig4}
\end{figure}

As discussed in the preceding section, there is a clear dependence of dielectric constant of the superconducting material (Eq.~\ref{eq-3}) on the operating temperature. In this regard, in this section, we address the influence of the operating temperature on the transmission spectra Dodecannaci superconductor-metamaterial photonic quasicrystal. To evaluate the thickness of the superconductor dependent transmittance spectrum from multilayered stack arranged according to Dodecannaci sequence at generation number, $x= 3$ (a), $x= 4$ (b), $x= 5$ (c), and  $x= 7$ (d) and plotted in Fig.~\ref{fig4}.  Here, excluding temperature, all other parameters that are used to compute the transmission spectra in Fig.~\ref{fig1} are not changed. We observed a significant change in the transmission characteristics due to the variation of $T$. The results have clearly shown that the increase in the operating temperature leads to a redshift the transmission response. Moreover, the degree of redshift also increases with an increase in temperature. The rate of red shift may be because of the decrease in the number of the super electrons about the number of the normal electrons. 

\section{Conclusions}\label{Conclusions}

In this work, we presented a general theory for the propagation of electromagnetic waves in one-dimensional superconductor-metamaterial superlattice, which is fabricated in a quasiperiodic fashion in accordance with the generalized Dodecannaci recurrence relation. It is shown that the transmission response can be tuned efficiently by the operating temperature as well as by the thicknesses of the constituent materials. When the temperature or the thickness of the DNG metamaterial layer is increased, the transmission response is blue shifted, whereas the transmission peaks become wider. A low rate of blue shift is observed in case of change in thickness of the superconducting layers with narrowing the transmission peaks. The influence of temperature on Dodecannaci superconductor-metamaterial quasicrystal is also discussed and found that the transmission response is red shifted and response is not linear. Our simulation results show that the proposed one-dimensional Dodecannaci superconductor-metamaterial superlattice would be a promising device with satisfying performance to work as tunable perfect narrowband filters and may have many other potential applications.

\section*{Acknowledgements} The author acknowledges the HOD, Department of Electronics and Communication Engineering, the Director of Engineering and Technology, and the Vice-Chancellor, SRM Institute of Science and Technology, Chennai, for their continuous encouragement.


\begin{thebibliography}{99}
\bibitem{S_john_1987} S. John, Phys. Rev. Lett.  \textbf{58} (1987) 2486.
\bibitem{J_E_1987}	J. E. Yablonovitch, Phys. Rev. Lett.  \textbf{58} (1987) 2059.
\bibitem{K_Busch_2001} K. Busch, S. John, Tunable photonic crystals, In: C. M. Soukoulis (Eds.), Photonic Crystals and Light Localization in the 21st Century, Springer, Dordrecht (2001) 41.
\bibitem{J_Y_2008}J. Y. Guo, H. Chen, H. Q. Li, Y. W. Zhang, Chin. Phys. B \textbf{17} (2008) 2544.
\bibitem{A_Aghajamali_2012} A. Aghajamali, M. Barati, Physica B\textbf{407} (2012) 1287.
\bibitem{A_Aghajamali_2016} A. Aghajamali, Appl. Opt.\textbf{55} (2016) 6336.
\bibitem{J_M_2002}	J.-M. Lourtioz, A. De Lustrac, C. R. Phys. \textbf{3} (2002) 79.
\bibitem{Z_Li_2018} Z. Li, Z. Ge, X.Y. Zhang, Z.Y. Hu, D. Zhao, J.W. Wu, Indian J. Phys.(\textbf{93} (2018) 511.
\bibitem{A_Agha_2016_1} A. Aghajamali, C. J. Wu, Appl. Opt.\textbf{55} (2016) 2086.
\bibitem{S_k_2016} S. K. Srivastava, A. Aghajamali, Physica B \textbf{489} (2016) 67.
\bibitem {M_Kamp_2004}	M. Kamp, T. Happ, S. Mahnkopf, G. Duan, S. Anand, A. Forchel, Physica E \textbf{21} (2004) 802.
\bibitem {H_Hojo_2004}	H. Hojo, A. Mase, J. Plasma Fusion Res. \textbf{80} (2004) 89. 
\bibitem {C_Nayak_2018_1}	C. Nayak, A. Aghajamali, D. P. Patil, Indian J Phys. \textbf{93} (2018) 401.
\bibitem {C_Nayak_2019_1}	C. Nayak, C. H. Costa, A. Aghajamali, IEEE Trans. Plasma Sci. \textbf{47} (2019) 1726.
\bibitem {C_Nayak_2017_1} C. Nayak, A. Aghajamali, F. Scotognella, A. Saha, Opt. Mater. \textbf{72} (2017) 25.
\bibitem {C_Nayak_2017_2}	C. Nayak, A. Aghajamali, A. Saha, Superlattice Microstruct. \textbf{111} (2017) 248. 
\bibitem {C_Nayak_2019_2}	C. Nayak, A. Aghajamali, A. Saha, N Das, Int. J. Mod. Phys. B \textbf{33} (2019), 1950219.
\bibitem {S_K_2017}	S. K. Srivastava, A. Aghajamali, J. Supercond. Nov. Magn. \textbf{30} (2017) 343.
\bibitem {M_Zamani_2016}	M. Zamani, Physica C \textbf{520} (2016) 42.
\bibitem {A_Agha_2016_2} A. Aghajamali, Appl. Opt. \textbf{55} (2016) 9797.
\bibitem {A_H_2013}	A. H. Aly, J. Supercond. Nov. Magn. \textbf{ 26} (2013) 553.
\bibitem {S_K_2014}	S. K. Srivastava, J. Supercond. Nov. Magn. \textbf{27} (2014) 101.
\bibitem {J_Barv_2015}	J. Barvestani, Physica B \textbf{457} (2015) 218.
\bibitem {V_G_2016}	V. G. Veselago, Sov. Phys. Usp. \textbf{10} (1968) 509.
\bibitem {A_H_2016}	A. H Alya, A. Aghajamali, H. A. Elsayed, M. Mobarak, Physica C \textbf{528} (2016) 5.
\bibitem {M_Kohmoto_1987}	M. Kohmoto, B. Sutherland, K. Iguchi, Phys. Rev. Lett. \textbf{58} (1987) 2436.
\bibitem {C_H_2017}	C. H. Costa, M. S. Vasconcelos, U. L. Fulco, E. L. Albuquerque, Opt. Mater. \textbf{72} (2017) 756. 
\bibitem {H_Rahimi_2019}	H. Rahimi, Opt. Mater. \textbf{57} (2016) 264. 
\bibitem {M_Zamani_2019}	M. Zamani, M. Amanollahi, M. Taraz, Opt. Mater. \textbf{88} (2019) 187.
\bibitem {C_nayak_2017_3}	C. Nayak, A. Aghajamali, T. Alamfard, A. Saha, Physica B \textbf{525} (2017) 41. 
\bibitem {M_Belli_2017}	M. Bellingeri, A. Chiasera, I. Kriegel, F. Scotognella, Opt. Mater. \textbf{72} (2017) 403.
\bibitem {Z_V_2013}	Z. V. Vardeny, A. Nahata, A. Agrawal, Nat. Photonics \textbf{7} (2013) 177.
\bibitem {K_Edagawa_2014}	K. Edagawa, Sci. Technol. Adv. Mater. \textbf{15}  (2014) 034805.
\bibitem {Y_Trabelsi_2019}	Y. Trabelsi, N. Ben Ali, A. Elhawil, R. Krishnamurthy, M. Kanzari, I.S. Amiri, P. Yupapin, Results Phys. \textbf{13} (2019) 1.
\bibitem {R_Talebzadeh_2018}	R. Talebzadeh, M. Bavaghar, Physica C \textbf{548} (2018) 119.
\bibitem {R_Luck_1993}	R. Lück, Int. J. Mod. Phys. B \textbf{7} (1993) 1437.
\bibitem {E_F_2019}	E. F. Silva, M. S. Vasconcelos, C. H. Costa, D. H. A. L. Anselmo, V. D. Mello, Opt. Mater. \textbf{98} (2019) 109450.
\bibitem {C_nayak_2020}	C. Nayak, Opt. Mater. \textbf{100} (2020) 109653.
\bibitem {M_Born_2013}	M. Born, E. Wolf, Principles of Optics, Cambridge University Press, Cambridge, (2013).
\bibitem {W_Buckel_2004}	W. Buckel, R. Kleiner, Superconductivity: Fundamentals and Applications, WILEY-VCH Verlag GmbH \& Co. KGaA, Weinheim, (2004).
\bibitem {A_Agha_2014_4}	A. Aghajamali, T. Alamfard, M. Barati, Physica B \textbf{454} (2014) 170.
\bibitem {A_Agha_2013}	A. Aghajamali, M. Hayati, C. J. Wu, M. Barati, J. Electromagn. Waves Appl. \textbf{27} (2013) 2317.
\bibitem {A_Agha_2014_5}	A. Aghajamali, B. Javanmardi, M. Barati, C. J. Wu, Optik \textbf{125} (2014) 839.
\bibitem {A_Agha_2015_4}	A. Aghajamali, T. Alamfard, M. Hayati, Optik \textbf{126} (2015) 3158.
\end{thebibliography}
\end{document}